\documentstyle[11pt,amsfonts]{article}

\newcommand{\be}{\begin{equation}}
\newcommand{\ee}{\end{equation}}
\newcommand{\bea}{\begin{array}}
\newcommand{\ea}{\end{array}}
\newcommand{\beqa}{\begin{eqnarray}}
\newcommand{\eeqa}{\end{eqnarray}}
\newcommand{\bean}{\begin{eqnarray*}}
\newcommand{\eean}{\end{eqnarray*}}

\def\starwedge{\stackrel{*}{\wedge}}

\def\up#1{\leavevmode \raise.16ex\hbox{#1}}

\newcommand{\gapproxeq}{\lower
 .7ex\hbox{$\;\stackrel{\textstyle >}{\sim}\;$}}
\newcommand{\lapproxeq}{\lower .7ex\hbox{$\;\stackrel
{\textstyle <}{\sim}\;$}}

\newcounter{appendice}

\def\thebibliography#1{{\bf REFERENCES\markboth
 {REFERENCES}{REFERENCES}}\list
 {[\arabic{enumi}]}{\settowidth\labelwidth{[#1]}\leftmargin\labelwidth
 \advance\leftmargin\labelsep
 \usecounter{enumi}}
 \def\newblock{\hskip .11em plus .33em minus -.07em}
 \sloppy
 \sfcode`\.=1000\relax}

\begin{document}
\centerline{ \LARGE Noncommutative Corrections to the Robertson-Walker metric}

\vskip 2cm

\centerline{ {\sc  S. Fabi, B. Harms and A. Stern  }  }

\vskip 1cm
\begin{center}

{\it Department of Physics, University of Alabama,\\
Tuscaloosa, Alabama 35487, USA}

\end{center}

\vskip 2cm

\vspace*{5mm}

\normalsize
\centerline{\bf ABSTRACT} 

Upon applying Chamseddine's  noncommutative deformation of gravity we obtain the leading order noncommutative corrections to the Robertson-Walker metric tensor.  We get an isotropic inhomogeneous metric tensor  for a certain choice of the noncommutativity parameters.  Moreover, the singularity  of the commutative metric at $t=0$ is replaced by a more involved space-time structure in the noncommutative theory.  In a toy model we  construct a scenario where there is no singularity at $t=0$ at leading order in the noncommutativity parameter. Although  singularities may still be present for nonzero $t$, they need not be the source of all time-like geodesics and the result resembles a bouncing cosmology.  

\vspace*{5mm}

\newpage
\scrollmode

\section{Introduction}

\setcounter{equation}{0}

Noncommutative deformations of general relativity  offer the promise of modeling effects of quantum gravity.  A number of different deformations have been given.\cite{Chamseddine:2000si},\cite{Aschieri:2005yw},\cite{Calmet:2006iz},\cite{Banerjee:2007th}  The approach of Aschieri et. al.\cite{Aschieri:2005yw} has the advantage of preserving the full diffeomorphism symmetry of the commutative theory.  As it is technically rather involved it so far however has not been very convenient for practical applications.  An older approach of  Chamseddine\cite{Chamseddine:2000si} is based on the noncommutative analogue $SO(4,1)$ gauge theory via the Seiberg-Witten map\cite{Seiberg:1999vs}.  It makes contact with general relativity using a Wigner-In\"on\"u contraction.  Ideally, one could then look for solutions to a noncommutative deformation of the field equations and map back to the commutative theory in order to obtain a physical interpretation.  This procedure could be easily  carried out in the case of $U(1)$ gauge theory in order  to obtain noncommutative  corrections to the Coulomb solution.\cite{Stern:2007an}
  However in the case of gravity, a deformation of the Einstein equations which is covariant under a noncommutative version of local Lorentz transformations remains obscure within the $SO(4,1)$ gauge theory approach.  An alternative procedure has been adopted  recently to obtain noncommutative corrections to black holes.\cite{Chaichian:2007we},\cite{Mukherjee:2007fa} (See also,\cite{Dolan:2006hv},\cite{Kobakhidze:2007jn},\cite{Banerjee:2008gc},\cite{Buric:2008th},\cite{Nicolini:2008aj}.)
 There, rather than solving some noncommutative analogue of the Einstein equations subject to the appropriate boundary conditions, one maps the known black hole solutions of general relativity to the noncommutative theory  and the defines a noncommutative analogue of the metric tensor in order to interpret the results.  As is typical with noncommutative gravity the leading order corrections are second order in the noncommutativety parameter.\cite{Mukherjee:2006nd}

Cosmology offers another possible realm of application of noncommutativity.
  Previous studies have led to corrections to the cosmic microwave background radiation\cite{Akofor:2007fv}, and noncommutative scalars fields have been coupled to the Robertson-Walker metric tensor in order to study effects on inflation.\cite{Lizzi:2002ib},\cite{Brandenberger:2002nq},\cite{Bertolami:2002eq}  Noncommutativity could also potentially resolve the big bang singularity.
  Here we apply the  procedure discussed above to obtain leading order corrections to the Robertson-Walker metric tensor. We get an isotropic inhomogenous metric tensor (with respect to one world line) after making a specific choice of the noncommutativity parameters. Isotropic inhomogenous cosmologies have been studied previously\cite{kras}, and some specific models have been proposed to look for 
explanations of the  cosmological acceleration.\cite{Rasanen:2003fy},\cite{Kolb:2004am},\cite{Buchert:2005xf},\cite{Nambu:2005zn},\cite{Romano:2006yc},
\cite{Enqvist:2007vb},
\cite{Jakacka:2008gd}  For an arbitrary expansion, the second order corrections to the Robertson-Walker metric tensor which we obtain are rather involved.  They simplify considerably for the special case of a linear expansion which allows for an analysis at small time $t$ (associated with the noncommutativity scale).   In this toy model   we  can construct a scenario where the noncommutative  metric tensor is everywhere  well defined at $t=0$  to leading order in the noncommutativity scale.  New singularities do appear at nonzero $t$ in this case, but  these  singularities are not the source of all time-like geodesics.  Instead, geodesics can be extended  through the $t=0$ time slice, and range  from  $t\rightarrow -\infty$ to $t\rightarrow +\infty$. The noncommutative  metric tensor is invariant under $t\rightarrow -t$ and describes a bouncing universe. 
 
 We review the gauge theory formalism for gravity in section 2 and the noncommutative generalization obtained by Chamseddine in section 3.  There we introduce a recursion relation found  recently in \cite{Ulker:2007fm} 
 for the second  order potentials.  It is employed in obtaining  the leading noncommutative corrections to the Robertson-Walker metric in section 4.   There we analyze the resulting space-time structure near $t=0$ for the case of  a linear expansion.  We  briefly remark on a slightly  more realistic expansion associated with a flat radiation dominated universe in section 5.  
 
\section{Commutative theory}

\setcounter{equation}{0}

The gauge theory formalism for gravity\cite{Chamseddine:2005td}
  is expressed in terms of spin connection and vierbein  one-forms,  $\omega^{ab}=-\omega^{ba}$ and $e^a$,  respectively.    $a,b,...=0,1,2,3$ are Lorentz indices which are raised and lowered with the flat metric tensor $\eta={\rm diag} (-1,1,1,1)$, while the space-time metric is
\be g_{\mu\nu}= e^a_{\;\;\;\mu}e^b_{\;\;\;\nu}\eta_{ab} \;.\label{mtrcnendom} \ee
 Infinitesimal variations of   $\omega^{ab}$  and $e^a$ induced from  local $ISO(3,1)$ transformations
are given by 
\beqa
 \delta \omega^{ab} &=& d\lambda ^{ab} +[\omega,\lambda]^{ab} \cr & &\cr
 \delta e^{a} &=& d\rho^{a} +  \omega^a_{\;\;\;c}\rho^{c} - \lambda ^a_{\;\;\;c} e^{c} \;,\label{cmtvgtfeo}
  \eeqa
for infinitesimal parameters   $\lambda ^{ab}=-\lambda ^{ba}$ and $\rho^a$, and where $[\omega,\lambda]^{ab}= \omega^a_{\;\;\;c}\lambda ^{cb} - \lambda^a_{\;\;\;c}\omega ^{cb}$.  The spin curvature and torsion two-forms, $R^{ab}=-R^{ba}$ and $T^a$ , respectively, are constructed from  $\omega^{ab}$ and $e^a$  according to 
\beqa  R^{ab}&=&d\omega^{ab} +
\omega^a_{\;\;\;c}\wedge\omega^{cb} \cr & &\cr
T^a&=&de^a
+\omega^a_{\;\;\;b} \wedge e^b \;,\eeqa
 and they satisfy the Bianchi identities
 \beqa  dR^{ab}&=&
R^a_{\;\;\;c}\wedge\omega^{cb}-
R^b_{\;\;\;c}\wedge\omega^{ca} \cr & & \cr
dT^a&=&R^a_{\;\;\;b} \wedge e^b -\omega^a_{\;\;\;c}\wedge T^{c}\;.\eeqa
The  field action
\be S=\frac 14 \int \epsilon_{abcd} R^{ab}\wedge e^c\wedge e^d \label{pltniactn}\;,\ee  describing pure gravity 
is invariant under local Lorentz transformations (and the full set of local Poincar\'e transformations when the torsion vanishes).  The field equations obtained from arbitrary variations of $\omega^{ab}$ and $e^a$ are 
 \beqa   
  T^{[a }\wedge e^{b]} &=&0  \label{mplztrsn} \\ & &\cr
          R^{[ab}\wedge e^{c]} &=&0\;,\label{cmtvpe} 
          \eeqa
 where the brackets indicate antisymmetrization of indices.  Provided that the vierbeins have an inverse, (\ref{mplztrsn}) implies a vanishing torsion,   while  (\ref{cmtvpe}) implies a vanishing Ricci curvature ${\cal R}_{\mu\nu}= 
{\cal R}_{\mu\sigma\nu}^{\quad \;\;\sigma}$, where the Riemann curvature  ${\cal R}_{\mu\nu\rho}^{\;\;\quad \lambda }$ is given in terms of the spin curvature by
\be {\cal R}_{\mu\nu\rho}^{\;\;\quad \lambda } = -R^{ab}_{\mu\nu} e_{b\rho} [e^{-1}]_{\;\;a}^\lambda  \;,\label{riemann} \ee
where $e^b_{\;\;\rho}[e^{-1}]^{\rho}_{\;\;a} =\delta^b_a$.

The above $ISO(3,1)$ gauge theory is obtained from a Wigner-In\"on\"u contraction of $SO(4,1)$ gauge theory.  Denote the potential one forms and the infinitesimal gauge parameters of $SO(4,1)$ gauge theory by $A^{AB}=-A^{BA}$ and $\Lambda^{AB}=-\Lambda^{BA}$, respectively, with indices $A,B,...=0,1,2,3,4$ which are raised and lowered with the metric tensor ${\rm diag} (-1,1,1,1,1)$.  An $SO(4,1)$ gauge variation is given by
\be  \delta A^{AB} = D\Lambda^{AB}= d\Lambda^{AB} + [A,\Lambda]^{AB} \;,\label{cmtvgt}\ee where $[A,\Lambda]^{AB}= A^A_{\;\;\;C}\Lambda^{CB} -  \Lambda ^A_{\;\;\;C}A^{CB} $, and
the curvature two forms $F^{AB}=-F^{BA}$  are
\be F^{AB}=dA^{AB} +
A^A_{\;\;\;C}\wedge A^{CB}\;. \ee
The contraction is obtained by setting 
\beqa \Lambda^{ab} =\lambda ^{ab} &\qquad & \Lambda^{a4} = \kappa \rho^a \cr & &\cr
A^{ab} =\omega^{ab} &\qquad & A^{a4} = \kappa e^a \cr & &\cr
 F^{ab} =R^{ab} &\qquad & F^{a4} = \kappa T^a \;,\eeqa
and taking the limit $\kappa\rightarrow 0$.

\section{Noncommutative theory}

\setcounter{equation}{0}

The noncommutative generalization for gauge theories based on non unitary groups was obtained in  \cite{Jurco:2000ja},\cite{Bonora:2000td}.
For the case of   $SO(4,1)$ gauge theory, denote  by $\hat A^{AB}$ and $\hat \Lambda^{AB}$, respectively,
 the noncommutative  analogues of the $SO(4,1)$ connection one forms and infinitesimal gauge parameters.  
 The noncommutative analogue of (\ref{cmtvgt})  is given by
\be  \delta \hat A^{AB} = D_\star\hat \Lambda^{AB}= d\hat \Lambda^{AB} + [\hat A,\hat \Lambda]^{AB}_\star \;,\label{ncgvrtns}\ee 
where $[\hat A,\hat \Lambda]^{AB}_\star =\hat A^A_{\;\;\;C}\star\hat \Lambda^{CB} - \hat \Lambda^A_{\;\;\;C}\star\hat A^{CB}\;,$ and the $\star$ denotes the Groenewold-Moyal star product.  Acting between two functions the latter  is given by
\be \star = \exp\;\biggl\{ \frac {i}2 \Theta^{\mu\nu}\overleftarrow{
  \partial_\mu}\;\overrightarrow{ \partial_\nu} \biggr\} \;,\label{gmstr} \ee  where
 $\Theta^{\mu\nu}=-\Theta^{\nu\mu}$ are constant matrix elements denoting the noncommutativity parameters and
  $\overleftarrow{
  \partial_\mu}$ and $\overrightarrow{ \partial_\mu}$ are left and right
derivatives,
respectively, with respect to  some coordinates $x^\mu$.   The noncommutative  analogue of the  $SO(4,1)$ curvature two form is 
\be \hat F^{AB}=d\hat A^{AB} +
\hat A^A_{\;\;\;C}\starwedge \hat A^{CB}\;. \ee
   $\starwedge$ denotes an exterior product   where
the usual pointwise product  between components of the
forms replaced by the Groenewold-Moyal star product.
The noncommutative spin connection,  vierbein, curvature and torsion forms, denoted respectively  by  $\hat \omega^{ab}$, $\hat e^a$, $\hat R^{ab}$ and $\hat T^a$ can be extracted from  $\hat A^{AB}$ as in the commutative case, i.e.,
 \beqa 
\hat A^{ab} =\hat \omega^{ab} &\qquad & \hat A^{a4} = \kappa \hat e^a \cr & &\cr
 \hat F^{ab} =\hat R^{ab} &\qquad & \hat F^{a4} = \kappa \hat T^a \;, \qquad {\rm as}\; \kappa\rightarrow 0\;.\label{ncvrbnscns}\eeqa

It is  known\cite{Jurco:2000ja},\cite{Bonora:2000td} that $\hat A$, $\hat F$ and $\hat \Lambda$, unlike their commutative analogues, are not valued in the $SO(4,1)$ Lie algebra, since  $(\hat A^{AB},\hat \Lambda^{AB})\rightarrow (-\hat A^{BA},-\hat \Lambda^{BA})$ is not an isomorphism of the gauge algebra (\ref{ncgvrtns}).  Moreover, $\hat A^{AB}, \hat F^{AB}$ and $ \hat \Lambda^{AB}$ cannot be restricted to real-valued forms, although one can impose antihermiticity
\beqa 
\hat A^{AB\;^*} &=&-\hat A^{BA} \cr & &\cr
\hat F^{AB\;^*} &=&-\hat F^{BA} \cr & &\cr
\hat \Lambda^{AB\;^*} &=&-\hat \Lambda^{BA}\;,\label{hrmtct} \eeqa and the 
diagonal components are purely imaginary.
It was observed\cite{Bonora:2000td}
 that if one enlarges the domain of  $\hat A^{AB}, \hat F^{AB}$ and $ \hat \Lambda^{AB}$ to the product of the space-time manifold (coordinatized by $x^\mu$) with the space of all noncommutativity parameters $\Theta^{\mu\nu}$, then the following conditions 
can be imposed consistent with the gauge algebra:
\beqa 
\hat A^{AB}(x,\Theta) &=&-\hat A^{BA}(x,-\Theta) \cr & &\cr
\hat F^{AB}(x,\Theta) &=&-\hat F^{BA}(x,-\Theta) \cr & &\cr
\hat \Lambda^{AB}(x,\Theta) &=&-\hat \Lambda^{BA}(x,-\Theta)\label{otmrfsm}\;. \eeqa  $\hat A^{AB}(x,\Theta)$, $\hat F^{AB}(x,\Theta)$ and $\hat \Lambda^{AB}(x,\Theta)$ can be expanded in terms of a power series  in $\Theta^{\mu\nu}$
\beqa 
\hat A^{AB}_\mu (x,\Theta) &=& A^{AB}_\mu(x)\;+\;\left.\matrix{ \cr A^{AB}_{\mu}\cr{}^{ (1)}\cr}\right.(x)\;+\;\left.\matrix{ \cr A^{AB}_{\mu}\cr{}^{ (2)}\cr}\right.(x)\;+\;\cdots\cr & &\cr
\hat F^{AB}_{\mu\nu}(x,\Theta) &=& F^{AB}_{\mu\nu}(x)\;+\;\left.\matrix{ \cr F^{AB}_{\mu\nu}\cr{}^{ (1)}\cr}\right.(x)\;+\;\left.\matrix{ \cr F^{AB}_{\mu\nu}\cr{}^{ (2)}\cr}\right.(x)\;+\;\cdots\cr & &\cr
\hat \Lambda^{AB}(x,\Theta) &=& \Lambda^{AB}(x)\;+\;\left.\matrix{ \cr \Lambda^{AB}\cr{}^{ (1)}\cr}\right.(x)\;+\;\left.\matrix{ \cr \Lambda^{AB}\cr{}^{ (2)}\cr}\right.(x)\;+\;\cdots\;,
 \label{pwrsrsxpnht}\eeqa where the $(n)$ subscript indicates the $n$th order in  $\Theta^{\mu\nu}$,
 \be  \left.\matrix{ \cr M^{AB}\cr{}^{ (n)}\cr}\right.(x)  = M^{AB}_{\rho_1\sigma_1 \rho_2\sigma_2 \cdots \rho_n\sigma_n}(x)\Theta^{\rho_1\sigma_1}\Theta^{\rho_2\sigma_2}\cdots\Theta^{\rho_n\sigma_n}\;.
 \ee
Then (\ref{otmrfsm}) implies that the coefficients $ M^{AB}_{\rho_1\sigma_1 \rho_2\sigma_2 \cdots \rho_n\sigma_n}(x)$  are (anti) symmetric under interchange of the $A$ and $B$ indices for $n$ odd (even).  (\ref{hrmtct}) then implies that the coefficients are imaginary (real) for $n$ odd (even). 

 The power series (\ref{pwrsrsxpnht}) have been defined using the Seiberg-Witten map from the commutative gauge theory\cite{Jurco:2000ja},\cite{Bonora:2000td} 
\be \hat A_\mu = \hat A_\mu(A) \qquad \hat F_{\mu\nu} = \hat F_{\mu\nu}(A) \qquad\hat \Lambda= \hat \Lambda(A,\Lambda) \;,\ee where 
$ A$, $  F$ and $  \Lambda$ again denote the commutative potentials, curvatures and infinitesimal  gauge parameters, respectively.  Since the latter are valued in the $SO(4,1)$ Lie algebra, this puts restrictions on the allowable   $ \hat A,  \hat F$ and $ \hat \Lambda$.  The  Seiberg-Witten map\cite{Seiberg:1999vs} then defines the space $\hat{ \cal A}$ of allowable noncommutative  potentials $ \hat A$.  
The map  is required to satisfy
\be \hat A_\mu ( A + \partial\Lambda +[A,\Lambda])-\hat A_\mu(A)  =\partial_\mu\hat\Lambda(\Lambda,A)
+[\hat A_\mu(A), \hat \Lambda(\Lambda,A)]_\star \;, \ee  for infinitesimal  $\Lambda$.
The zeroth order in the expansion (\ref{pwrsrsxpnht}) agrees with the commutative theory. Up to homogeneous terms, the first order expressions for the noncommutative potentials and infinitesimal gauge parameters are 
\beqa 
  \left.\matrix{ \cr A_{\mu}\cr{}^{ (1)}\cr}\right. &=&-\frac i4 \Theta^{\rho\sigma} \{ A_\rho ,\partial_\sigma A_\mu+ F_{\sigma\mu}\}\cr & &\cr \left.\matrix{ \cr \Lambda\cr{}^{ (1)}\cr}\right.
 &=&-\frac i4\Theta^{\rho\sigma} \{   A_\rho,\partial_\sigma \Lambda \}\;,
 \label{frstrdrSW}\eeqa
  where  the parentheses denote the anticommutator $\{ A ,B\}^{AB} =A^{AC}B_C^{\;\;B} +B^{AC}A_C^{\;\;B}$. 
 Recently, a relatively simple recursion relation was found for the higher order potentials and gauge parameters.\cite{Ulker:2007fm}
  At second order one gets
\beqa 
  \left.\matrix{ \cr A_{\mu}\cr{}^{ (2)}\cr}\right. &=&-\frac i8 \Theta^{\rho\sigma} \biggl(\{ \left.\matrix{ \cr A_{\rho}\cr{}^{ (1)}\cr}\right. ,\partial_\sigma A_\mu+ F_{\sigma\mu}\}
 +\{ A_\rho ,\partial_\sigma  \left.\matrix{ \cr A_{\mu}\cr{}^{ (1)}\cr}\right.+  \left.\matrix{ \cr F_{\sigma\mu}\cr{}^{ (1)}\cr}\right.\}+\{ A_\rho ,\partial_\sigma A_\mu+ F_{\sigma\mu}\}_{\star_{(1)}}\biggr)\cr & &\cr \left.\matrix{ \cr \Lambda\cr{}^{ (2)}\cr}\right.
 &=&-\frac i8\Theta^{\rho\sigma}\biggl( \{ \left.\matrix{ \cr A_{\rho}\cr{}^{ (1)}\cr}\right., \partial_\sigma \Lambda \}+\{ A_\rho, \partial_\sigma\left.\matrix{ \cr \Lambda\cr{}^{ (1)}\cr}\right. \}+\{  A_\rho,\partial_\sigma \Lambda \}_{\star_{(1)}}\biggr)\;,
 \label{sostrdrSW}\eeqa
where the subscript $\star_{(n)}$ on the  bracket indicates the $n$th order term in the $\Theta$ expansion of the star-anticommutator $\{ A ,B\}_\star^{AB} =A^{AC}\star B_C^{\;\;B} +B^{AC}\star A_C^{\;\;B}$.

Using (\ref{ncvrbnscns})   one  next defines  noncommutative vierbeins and spin connections through a power series expansion in  $\Theta$\cite{Chamseddine:2000si}
\beqa 
\hat e^{a}_\mu (x,\Theta) &=& e^{a}_\mu(x)\;+\;\left.\matrix{ \cr e^{a}_{\mu}\cr{}^{ (1)}\cr}\right.(x)\;+\;\left.\matrix{ \cr e^{a}_{\mu}\cr{}^{ (2)}\cr}\right.(x)\;+\;\cdots
\cr & &\cr \hat \omega^{ab}_\mu (x,\Theta) &=& \omega^{ab}_\mu(x)\;+\;
\left.\matrix{ \cr \omega^{ab}_{\mu}\cr{}^{ (1)}\cr}\right.(x)\;+\;\left.\matrix{ \cr \omega^{ab}_{\mu}\cr{}^{ (2)}\cr}\right.(x)\;+\;\cdots\;,\label{xpnsfreom}\eeqa
which in turn is defined through the Seiberg-Witten map  of the potential one forms
\be \hat e =\hat e(e,\omega) \qquad \hat \omega =\hat\omega(e,\omega)\;. \ee  
The zeroth order again agrees with the commutative theory, while for the first and second orders one gets 
\beqa \left.\matrix{ \cr e^{a}_{\mu}\cr{}^{ (1)}\cr}\right. &=&-\frac i4 \Theta^{\rho\sigma}\biggl([ \omega_\rho]^{a}_{\;\;b} (\partial_\sigma e^b_{\mu}+ T^b_{\sigma\mu}) + (\partial_\sigma\omega_\mu + R_{\sigma\mu})^{a}_{\;\;b}e^b_{\rho}\biggr)
 \cr & &\cr \left.\matrix{ \cr \omega^{ab}_{\mu}\cr{}^{ (1)}\cr}\right. &=& -\frac i4 \Theta^{\rho\sigma} \{ \omega_\rho ,\partial_\sigma \omega_\mu+ R_{\sigma\mu}\}^{ab}\label{frstordreom}\;,\eeqa
 and
 \beqa \left.\matrix{ \cr e^{a}_{\mu}\cr{}^{ (2)}\cr}\right. &=& -\frac i8 \Theta^{\rho\sigma}\biggl([ \left.\matrix{ \cr \omega_{\rho}\cr{}^{ (1)}\cr}\right.]^{a}_{\;\;b} (\partial_\sigma e^b_{\mu}+ T^b_{\sigma\mu})+[\omega_\rho]^{a}_{\;\;b} (\partial_\sigma \left.\matrix{ \cr e^b_{\mu}\cr{}^{ (1)}\cr}\right.+ \left.\matrix{ \cr T^b_{\sigma\mu}\cr{}^{ (1)}\cr}\right.)+[\omega_\rho]^{a}_{\;\;b} \star_{(1)}(\partial_\sigma e^b_{\mu}+ T^b_{\sigma\mu})\cr & &\cr & & + (\partial_\sigma \left.\matrix{ \cr \omega_{\mu}\cr{}^{ (1)}\cr}\right. + \left.\matrix{ \cr R_{\sigma\mu}\cr{}^{ (1)}\cr}\right.)^{a}_{\;\;b}e^b_{\rho}+ (\partial_\sigma\omega_\mu + R_{\sigma\mu})^{a}_{\;\;b}\left.\matrix{ \cr e^b_{\rho}\cr{}^{ (1)}\cr}\right.+ (\partial_\sigma\omega_\mu + R_{\sigma\mu})^{a}_{\;\;b}\star_{(1)}e^b_{\rho}\biggr)\cr & &
 \cr & &\cr  \left.\matrix{ \cr \omega^{ab}_{\mu}\cr{}^{ (2)}\cr}\right. &=& -\frac i8 \Theta^{\rho\sigma} \biggl(\{ \left.\matrix{ \cr \omega_{\rho}\cr{}^{ (1)}\cr}\right. ,\partial_\sigma \omega_\mu+ R_{\sigma\mu}\}^{ab}
 +\{ \omega_\rho ,\partial_\sigma  \left.\matrix{ \cr \omega_{\mu}\cr{}^{ (1)}\cr}\right.+  \left.\matrix{ \cr R_{\sigma\mu}\cr{}^{ (1)}\cr}\right.\}^{ab}+\{ \omega_\rho ,\partial_\sigma \omega_\mu+ R_{\sigma\mu}\}_{\star_{(1)}}^{ab}\biggr)\;,\cr & &
 \label{soctns}
 \eeqa
 where the first order corrections to the curvature and torsion are defined as
 \beqa\left.\matrix{ \cr R^{ab}_{\mu\nu}\cr{}^{ (1)}\cr}\right. &=&\partial_\mu\left.\matrix{ \cr \omega^{ab}_{\nu}\cr{}^{ (1)}\cr}\right. -  \partial_\nu\left.\matrix{ \cr \omega^{ab}_{\mu}\cr{}^{ (1)}\cr}\right. +[\left.\matrix{ \cr \omega_{\mu}\cr{}^{ (1)}\cr}\right. ,\omega_\nu ]^{ab}+[\omega_\mu, \left.\matrix{ \cr \omega_{\nu}\cr{}^{ (1)}\cr}\right.  ]^{ab}+[\omega_\mu,  \omega_{\nu} ]_{\star_{(1)}}^{ab}\cr
 \left.\matrix{ \cr T^a_{\mu\nu}\cr{}^{ (1)}\cr}\right.&=& \partial_\mu\left.\matrix{ \cr e^{a}_{\nu}\cr{}^{ (1)}\cr}\right. + [ \left.\matrix{ \cr\omega_\mu]^{a}_{\;\;b} \cr{}^{ (1)}\cr}\right. e^b_{\nu} + [ \omega_\mu]^{a}_{\;\;b} \left.\matrix{ \cr e^b_{\nu}  \cr{}^{ (1)}\cr}\right. + [ \omega_\mu]^{a}_{\;\;b}\star_{(1)} e^b_{\nu}\; -\;(\mu \rightleftharpoons\nu)\;.
 \eeqa 
For the discussion below we  follow \cite{Chaichian:2007we},\cite{Mukherjee:2007fa} and  specialize to the case of zero torsion in the commutative theory; i.e.,
 \be T_{\mu\nu}^a=0 \;.\ee
 Furthermore, in order to make a physical interpretation of the noncommutative vierbeins we  define the real symmetric  noncommutative version of the metric tensor according to
 \be \hat g_{\mu\nu}=\frac 12 \eta_{ab} (\hat e ^a_{\mu} \star \hat e ^{b*}_{\nu} + \hat e ^b_{\nu} \star \hat e ^{a*}_{\mu})\;. \label{ncmtrctnsr}\ee

\section{Robertson-Walker metric}

\setcounter{equation}{0}

We now apply the above  formalism to the case of the Robertson-Walker metric.   Starting with the usual expression for the Robertson-Walker   invariant measure 
\be ds^2= -dt^2 + a(t)^2\Bigl(\frac{dr^2}{1-kr^2} +
 r^2(d\theta^2 + \sin^2 \theta \;d\phi^2)\Bigr)
\;,\ee where $a(t)$ is the scale factor,
one can assign vierbein one forms according to 
\be e^0 = dt \qquad e^1 = \frac {a(t)\; dr}{\sqrt{1-kr^2}} \qquad  e^2 = a(t) r\; d\theta\qquad  e^3 =a(t) r \sin\theta \;d\phi \;. \label{scwzvrbn}\ee  The torsion vanishes  upon choosing the following for the spin connection one forms
\be \left.\matrix{& & \cr \omega^{01} = \chi dr\;\qquad\qquad & \omega^{02} = \dot a r \;d\theta\qquad\qquad\quad & \omega^{03}= \dot a r \sin\theta \;d\phi\cr & & \cr \omega^{12} = -\sqrt{1-kr^2} \;d\theta &  \omega^{31} =\sqrt{1-kr^2} \sin\theta \;d\phi &  \omega^{23} =-\cos\theta\; d\phi \cr& & \cr}\right.\label{scwzspncn}\;,\ee 
where the dot denotes differentiation with respect to  $t$.   To determine $\chi$  one can compute the curvature scalar 
${\cal R}={\cal R}_{\mu\nu}^{\quad \;\mu\nu}$
using (\ref{riemann}),
and compare with the known result for the Robertson-Walker metric; i.e.,
\be {\cal R}=6\biggl(\frac {\ddot a}a + \Bigl( \frac {\dot a} a\Bigr)^2 + \frac k {a^2}\biggr)\;.\ee
  They agree for 
\be \chi = \frac{\dot a}{\sqrt{1-kr^2}}\;.\ee

For simplicity we set all components of $\Theta^{\mu\nu}$ equal to zero except for
\be \Theta^{tr}  = -\Theta^{rt}= \Theta \;.\label{stncvty}\ee  This choice leads to
 an isotropic inhomogeneous 
   space-time.  
Up to second order in $\Theta$, we find the following noncommutative vierbein one forms after substituting into
(\ref{xpnsfreom})-(\ref{soctns})
\beqa
\hat e^0 &=& dt+ \frac{i \Theta }4 \; \frac{\dot a^2+2 a\ddot a}{1-k r^2 }dr -\frac{5 \Theta ^2 \left(\ddot a^2+\dot a a^{(3)}\right)}{32 \left(1-kr^2 \right)} dt + \frac{r \Theta ^2 k  \left(9 \dot a\ddot a-2aa^{(3)}\right)}{16 \left(1-kr^2 \right)^2}dr
\cr & &\cr
    \hat e^1 &=& \frac{a\; dr}{\sqrt{1-k r^2}} + \frac{i \Theta \ddot adt}{4 \sqrt{1-kr^2 }}  -\frac{i r \Theta  k  \dot adr}{4 \left(1-kr^2  \right)^{3/2}}
    -\frac{3 \Theta ^2 \left(3 \ddot a\dot a^2+4 a a^{(3)} \dot a+4 a\ddot a^2\right)}{32 \left(1-kr^2 \right)^{3/2}} dr
  \cr & &\cr 
 \hat  e^2 &=& \Phi d\theta \qquad \qquad \cr & &\cr \hat  e^3 &=& \Phi \sin\theta d\phi\;,
   \eeqa
where 
\be \Phi= a r   -\frac{i \Theta }{4}  \dot a  - \frac{r \Theta ^2 \left(8 a\ddot a^2+\left(9 \dot a^2+4k \right)\ddot a+4a\dot a a^{(3)}\right)}{32 \left(1-kr^2\right)} \;,\ee and $a^{(3)}$ denotes the third time derivative of $a$.  
 Only one off diagonal metric tensor component (\ref{ncmtrctnsr}) results in these coordinates
 \beqa \hat g_{tt} &=& -1+ \frac{\Theta ^2 \left(6 \ddot a^2+5 \dot a a^{(3)}\right)}{16(1- kr^2)  } +O\left(\Theta ^4\right)
  \cr
 & &\cr \hat g_{rr} &=&    
    \frac{a^2}{1-kr^2  }-\frac{\Theta ^2 }{16 \left(1-kr^2\right)^3}\biggl(\left(1-kr^2 \right)\left(\dot a^4+13a\ddot a\dot a^2+12  a^2 a^{(3)} \dot a+16(a\ddot a)^2\right)\cr & &\qquad\qquad+k  \left(3 k  r^2+4\right)\dot a^2+4a\ddot a k (k  r^2+1)\biggr)+O\left(\Theta
   ^4\right) 
   \cr
 & &\cr \hat g_{\theta\theta} &=&  r^2 a^2+ \frac{\Theta ^2}{16}  \left(-\frac{a \left(8 a\ddot a^2+\left(9 \dot a^2+4 k \right)\ddot a+4 a\dot a  a^{(3)}\right) r^2}{1-kr^2}+5 \dot a^2+4 a\ddot a\right)+O\left(\Theta ^4\right)
 \cr
 & &\cr \hat g_{\phi\phi} &=&    \sin ^2\theta\;\hat g_{\theta\theta}
    \cr
 & &\cr  \hat g_{tr} &=& -\frac{r \Theta ^2 k\dot a\ddot a}{2 \left(1-kr^2 \right)^2}  +O\left(\Theta ^4\right)\;.
\label{trncmtrctnsr}
 \eeqa
When treated as a standard  metric tensor it is associated with an inhomogeneous isotropic space-time with respect to the worldline at $r=0$.\footnote{This is not the case  for generic $\Theta^{\mu\nu}$.}
 We note that there are no second order corrections when the scale factor is a constant.
 
 We wish to examine the noncommutative  metric tensor for small $t$ (which we define later).  As a general analysis with arbitrary scale factor is quite involved, we shall  examine a toy model.  The simplest nontrivial example is the case of  $a(t)=vt$, associated with a linear expansion in the commutative theory.\footnote{If one further restricts $k=-v^2$, then the commutative theory corresponds to the Milne universe.  In this case, all components of the Riemann curvature vanish and the commutative metric can be mapped into a region of Minkowski space using $(t,r)\rightarrow (t'=  t\sqrt{1+v^2r^2}\;,\;r'= vtr)$.}     Here we can construct a scenario where there is no singularity for $t=0$ at leading order in $\Theta$.  We first note that the case of  $a(t)=vt$ implies that  the off diagonal matrix element $\hat g_{tr}$ vanishes at leading order and that the diagonal elements are invariant under $t\rightarrow -t$. 
 If in the noncommutative theory we  define the analogue of the invariant measure according to $d\hat s^2=\hat g_{\mu\nu}dx^\mu dx^\nu$, it here  has the form
 \beqa d\hat s^2&=& -dt^2 +  \frac{ a_r(t,r)^2\; dr^2}{1-kr^2}
 + r^2 a_\Omega(t,r)^2\;(d\theta^2+\sin\theta^2d\phi^2)+O\left(\Theta ^4\right) \label{inhomoiso}
\;,\eeqa
 where  
\beqa a_r(t,r)^2 &=& a(t)^2 -\frac{\Theta ^2  v^2}{16}\biggl( \frac{ v^2}{1-kr^2} +\frac{k(3 k  r^2+4)}{ (1-kr^2 )^2}\biggl)\cr
a_\Omega(t,r)^2 &=&  a(t)^2+\frac{5 \Theta ^2 v^2}{16r^2}\;. \eeqa 
The second order correction to $
a_\Omega(t,r)$ renders $\hat g_{\theta\theta}$ and $  \hat g_{\phi\phi}$ nonsingular at $t=0$. The second order correction to   $a_r(t,r)^2$ is   everywhere negative when 
\be -\frac{v^2}4<k\le 0\;,\label{vsqovrfr}\ee
 which means that  $\hat g_{rr}$ is also everywhere nonsingular at $t=0$. Thus when (\ref{vsqovrfr}) holds, the leading corrections imply that there is no singularity at $t=0$.  Instead,   the noncommutative  metric tensor is everywhere  well defined on the $t=0$ time slice, which has a three-dimensional Minkowski  signature $(-1,1,1)$. The same result applies for
\be k>0\;,\qquad 0\le r^2 < \frac 1k \;.\label{vsqovrfrkps}\ee  
(The metric tensor is ill-defined at $r^2=1/k$ for the case of $k>0$.)
  On the other hand, the noncommutative metric tensor is singular for these two cases when
\be t^2 =\frac{\Theta ^2  }{16}\biggl( \frac{ v^2}{1-kr^2} +\frac{k(3 k  r^2+4)}{ (1-kr^2 )^2}\biggl)\label{hrzn}\;,\ee
the solutions of which define two disconnected surfaces, associated with positive and negative values for $t$. [For the choice of a dimensionful radial coordinate, $\Theta^{-1}$, $v^2$ and $k$ have units of 1/length${}^2$.]  One can compute the scalar curvature in order to determine whether or not the surfaces correspond to  coordinate singularities. Treated as an ordinary space-time metric tensor, $\hat g_{\mu\nu}$ leads to the following (commutative) space-time scalar curvature\footnote{ Alternatively, one can define a noncommutative analogue of the scalar curvature, as is done in \cite{Mukherjee:2007fa}, however its geometrical meaning is not obvious.}
\beqa {\cal R} &=&
{\cal R}_{\mu\nu}^{\quad \;\mu\nu}=\frac{6 \left(v^2+k \right)}{t^2 v^2}\cr & &\cr & &-\frac{\Theta ^2 \left(k  \left(v^4+8k  v^2+7k
   ^2\right) r^6-\left(v^4+26 k v^2+2 k ^2\right) r^4+\left(11 v^2+4 k \right) r^2+5\right)}{8
   r^4 t^4 v^2 \left(1-r^2k\right)^2}\cr & &\cr& &+O\left(\Theta^4\right)\;.\label{soscrvtr}
\eeqa
It is well behaved everywhere  on the surfaces  defined by (\ref{hrzn})  except at the spatial origin.   It follows from the second order analysis that there are (at least) two singular points on the space-time manifold,
\be (t,r)=(\pm\frac\Theta 4\sqrt{v^2+4k},0)\label{snglrpnt}\;,\label{tosnglrts}\ee
  which go to the big bang singularity  when $\Theta\rightarrow 0$.\footnote{The scalar curvature   given in (\ref{soscrvtr}) is still singular at $t=0$.  However, this is due to the truncation of the expansion in $\Theta$.  The exact expression for the scalar curvature which follows from the second order corrected metric tensor is well defined at $t=0$ for (\ref{vsqovrfr}) or (\ref{vsqovrfrkps}).}  [Eq. (\ref{tosnglrts}) can be used to define `small $t$' in this case.]   
   Unlike the big bang singularity, the two singular points in  (\ref{tosnglrts})  are not the source of all time-like geodesics when $\Theta\ne 0$.  
To see this we next look at the geodesic equations.  Call $u^\mu=\frac {dx^\mu}{d\sigma}$ where $\sigma$ parametrizes the geodesic.  Due to the rotational invariance we can consistently set $u^\theta=u^\phi=0$.  The geodesic equations for $u^t$ and $u^r$ then read
 \beqa \frac {du^t}{d\sigma}&=&-\frac{t (v u^r)^2}{1-kr^2}+O\left(\Theta ^4\right)\cr & &\cr
 \frac {du^r}{d\sigma}&=&-\frac{kr  (u^r)^2}{1-kr^2}-\frac{2 u^t u^r}{t}-\frac{\Theta ^2}{16 t^3 \left(kr^2 
   -1\right)^3}\biggl\{r t k  \Bigl(\left(1-kr^2
   \right) v^2+k  \left(3k  r^2+11\right)\Bigr) (u^r)^2\cr & &\cr& &\qquad -2 \left(1-kr^2 \right) \Bigl(\left(1-kr^2 \right)
   v^2+k  \left(3 k  r^2+4\right)\Bigr) u^t u^r\biggr\}\;+\;O\left(\Theta ^4\right)\;.
 \eeqa
 The  comoving world lines  $u^t=1,\;\;u^r=0$ of the commutative theory are unaffected by the second order corrections in $\Theta$.  Consequently, all of them, except for the central one at $r=0$ which intersects  the singular points (\ref{snglrpnt}), can be extended  through the $t=0$ time slice, and range  from  $t\rightarrow -\infty$ to $t\rightarrow +\infty$.   Therefore, although  cosmic singularities are still present at leading order in $\Theta$, they are no longer the source of all time-like geodesics.  This result   also holds when (\ref{vsqovrfr}) or (\ref{vsqovrfrkps}) are no longer true, as is the case with the Milne universe.  Then there are singularities on  the $t=0$ time-slice on the surface of a sphere of radius\footnote{The  arguement given in the previous footnote suggests that these are coordinate singularities.}  
 \be r= \sqrt{\frac{v^2+4k}{k(v^2-3k)}}\;,\ee 
 but they are not the source of all time-like geodesics.

\section{Concluding Remarks}

\setcounter{equation}{0}

It is  of course of interest to go beyond the toy model considered in the previous section and consider  
   more realistic   functions for the scale parameter.  Unfortunately, the analysis then becomes quite a bit more involved.
 For the example of $ a(t)=C t^{1/2}$, which is standardly associated with a flat radiation dominated universe, the noncommutative  metric tensor (\ref{trncmtrctnsr}) is  no longer diagonal in the coordinates $(t,r,\theta,\phi)$ unless $k =0$.   From (\ref{trncmtrctnsr}) we can compute  the volume form for this case
\be \det \hat g_{\mu\nu}=-\Biggl({ t^3  }-\frac{ \Theta ^2 \left(83 r^2 \left(1-kr^2\right) C^2-4t \left(-5 k^2 r^4+4 k  r^2+2\right)\right)  }{256r^2 \left(1-kr^2\right)^2}
   \Biggr)\;\frac{C^6r^4\sin
   ^2\theta} {1-kr^2 }\;.\label{detgmunu}\ee
It is well behaved at $t=0$ for $k\le 0$ and $k>0\;,\; 0\le r^2 < \frac 1k, $ except for the origin $r=0$.  
The origin appears to be a singularity in space-time from the expression for  the space-time scalar curvature which in this case is
$$ {\cal R} =
{\cal R}_{\mu\nu}^{\quad \;\mu\nu}=\frac{6 k }{C^2 t}\qquad\qquad\qquad\qquad \qquad\qquad\qquad\qquad\qquad\qquad \qquad\qquad\qquad$$ 
$$ +\frac{\left(933 C^4 \left(r^2k -1\right) r^4-4 C^2 t \left(82 k^2 r^4-284
   k r^2+21\right) r^2+16 t^2 \left(5 k^3 r^6-22 k^2 r^4-4 k r^2-1\right)\right) \Theta
   ^2}{512 C^2 r^4 t^5 \left(1-kr^2\right)^2}$$
   \be  +O\left(\Theta ^4\right)\;.\qquad\qquad\qquad\qquad \qquad\qquad\qquad\qquad \qquad\qquad\qquad\label{stctfrctoh}
\ee     
More generally, upon setting the parenthesis in (\ref{detgmunu})  equal to zero one now gets a cubic equation in $t$,  defining surfaces where the noncommuative metric tensor is singular. (\ref{stctfrctoh}) may be employed to determine whether or not points on these surfaces are coordinate singularities.  The geodesic equations for $u^t$ and $u^r$ now have  $\Theta^2$ terms proportional to $(u^t)^2$, and so unlike in the previous case, the  comoving world lines  $u^t=1,\;\;u^r=0$ of the commutative theory are not geodesics of the noncommutative metric due to $\Theta^2$ corrections. 
  
Of course it is also of interest to consider the example of $ a(t)=C t^{2/3}$, which is standardly associated with a matter dominated universe.  One can then try to perform spatial averages of the second order
correction in this case in order to obtain the best fit for the map of
the luminosity distance of the supernova versus the redshift along the lines of \cite{Rasanen:2003fy},\cite{Kolb:2004am},\cite{Buchert:2005xf},\cite{Nambu:2005zn},\cite{Romano:2006yc},
\cite{Enqvist:2007vb},\cite{Jakacka:2008gd}.  Since we are not required to make any particular choice for $\Theta^{\mu\nu}$, as in   (\ref{stncvty}),  a reasonable fit may be possible.

\end{document}